\documentclass[10pt, conference, letterpaper, final]{IEEEtran}
\IEEEoverridecommandlockouts
\usepackage{cite}
\usepackage{amsmath,amssymb,amsfonts}

\usepackage{algorithmicx}
\usepackage[noend]{algpseudocode}
\usepackage{graphicx}
\usepackage{textcomp}
\usepackage{xcolor}
\usepackage{subfig}
\usepackage{algorithm}
\usepackage{tabularx}
\usepackage{makecell}
\usepackage[colorinlistoftodos,prependcaption,textsize=tiny]{todonotes}
\usepackage[final]{changes}
\def\BibTeX{{\rm B\kern-.05em{\sc i\kern-.025em b}\kern-.08em
    T\kern-.1667em\lower.7ex\hbox{E}\kern-.125emX}}

\begin{document}

\title{Accelerating Handover in Mobile Satellite Network\\
 \thanks{This work has been supported by the National Natural Science Foundation of China Grant No. 62341105. The corresponding authors are Yue Gao and Xiong Wang. }
}

\author{Jiasheng Wu, Shaojie Su, Xiong Wang, Jingjing Zhang, Yue Gao  \\
Fudan University, China \\
\{22110240044, 22210240270\}@m.fudan.edu.cn, \{wangxiong, jingjingzhang, gao.yue\}@fudan.edu.cn}

\maketitle
\begin{abstract} 
The construction of Low Earth Orbit (LEO) satellite constellations has recently spurred tremendous attention from academia and industry. 5G and 6G standards have specified LEO satellite network as a key component of 5G and 6G networks. However, ground terminals experience frequent, high-latency handover incurred by satellites’ fast travelling speed, which deteriorates the performance of latency-sensitive applications. To address this challenge, we propose a novel handover flowchart for mobile satellite networks, which can considerably reduce the handover latency. The innovation behind this scheme is to
mitigate the interaction between the access and core networks that occupy the majority of time overhead by leveraging the predictable travelling trajectory and spatial distribution inherent in mobile satellite networks. Specifically, we design a fine-grained
synchronized algorithm to address the synchronization problem due to the lack of control signalling delivery between the access and core networks. Moreover, we minimize the computational complexity of the core network using information such as the satellite access strategy and unique spatial distribution, which is caused by frequent prediction operations. We have built a prototype for a mobile satellite network using modified Open5GS and UERANSIM, which is driven by actual LEO satellite constellations such as Starlink and Kuiper. We have conducted extensive experiments, and the results demonstrate that our proposed handover scheme can considerably reduce the handover latency compared to the 3GPP Non-terrestrial Networks (NTN) and two other existing handover schemes.

\end{abstract}


\section{Introduction}

The Internet service provided by Low earth orbit (LEO) satellites such as Starlink has been emerged rapidly \cite{starlink}. The satellite internet can provide global coverage and thus
is a valuable complement to traditional terrestrial network. To this end, 3GPP 5G and 6G standards have specified that satellite communication is a key component of the whole system to establish the mobile satellite network \cite{23501,23502,6Gwhite}, where several well-known telecommunication operators like T-Mobile collaborate with satellite network providers to provide world-wide communication services to users by directly building the link between satellites and mobile phones \cite{t-mobile,optus}.

\begin{figure}[t!]
\centering
   \subfloat[Transparent mode]{\includegraphics[width=\linewidth]{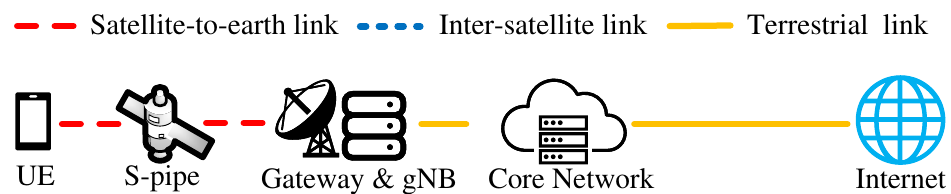}\label{pipe}}
   \quad
    \subfloat[Regenerative mode]{\includegraphics[width=\linewidth]{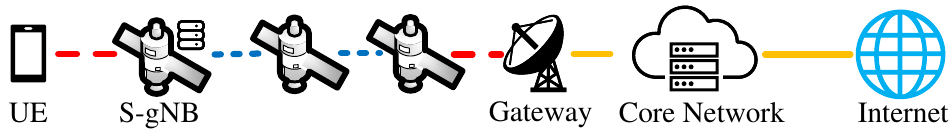}\label{regnerative}}
    \caption{Two different operating modes of mobile satellite network in (a) transparent mode and (b) regenerative mode.}
    \vspace{-0.6cm}
\end{figure}

There exist two operating modes in mobile satellite network, respectively as the transparent mode and the regenerative  mode \cite{38821}. Initially, most satellites adopt the transparent mode, as shown in Fig.~\ref{pipe}. In this case, satellites serve as transparent physical nodes between ground terminals. However, operating in this mode suffers from several obvious shortcomings such as limited coverage, single-point bottleneck, and relatively high latency \cite{stateless_mobile}. Recently, the regenerative  mode, where LEO satellites act as base stations and provide worldwide coverage by leveraging inter-satellite links (ISLs), has been proposed and attracts extensive attention, as described in Fig. \ref{regnerative}. Meanwhile, mobile satellite networks operating in the regenerative mode have been in test \cite{re-test}. In the following discussion, we focus on the regenerative mode and refer to these satellites as S-gNB (\textbf{S}atellite next \textbf{G}eneration \textbf{N}ode\textbf{B}).


However, there exists a critical challenge in the mobile satellite network, i.e., handovers are triggered frequently due to the fast travelling speed of satellites. Different from the terrestrial network, the S-gNB, as a crucial infrastructure for user access located at LEO satellite, travels at a high speed and is usually far away from the core network. Consequently, ground terminals experience handover between two satellites every 2-5 minutes, resulting in an average handover latency of around 400 ms. This high-latency handover declines the user experience, especially for the latency-sensitive applications.

Naturally, to deal with this challenge, one may wonder why not adopt the seamless handover strategy named soft handover designed in 3G \cite{soft-handover} and dual active protocol stack (DAPS) in 5G \cite{38300}. The core idea of above handover schemes is to support multiple parallel links with different base stations, which brings multiple times of hardware overhead \cite{9984697}. With regard to this factor, seamless handover is inappropriate and thus not applied in 5G communications. In mobile satellite network, maintaining multiple parallel links between the ground terminal and several satellites exacerbates the hardware overhead in order to mitigate the high channel loss (up to 160 dB) \cite{38811}, thus substantially increasing the system complexity.

In this paper, inspired by the concept of computing in the network \cite{Edge_Computing,6Gwhite}, we design a novel handover flowchart for mobile satellite network, which reduces handover latency through adding additional computation in the core network. The innovations behind the proposed scheme are to achieve handover only involving the interaction between the UE and access network (i.e., LEO satellites) based on the predictable travelling trajectory and unique spatial distribution of LEO satellites. In this way, interaction between the access and core networks that accounts for the majority of time overhead in the handover procedure is avoided since handover signaling transmission from the access network to core network should pass through multiple satellites and then to the ground. 

The implementation of the proposed handover scheme entails two main challenges. First, without control signaling interaction, it is non-trivial to maintain strict synchronization between the access and core networks because the core network has no access to the handover triggering time. Secondly, the proposed handover scheme requires predicting access satellites for all UEs based on the predicted trajectories, which imposes substantial computation overhead on the core network. To address the first challenge, we propose a fine-grained synchronized algorithm, where two specific time points are set to determine the access satellite of the UE. Meanwhile, we leverage the access strategy and spatial distribution features in LEO satellite networks to reduce the number of UEs and satellites required for prediction, thus considerably alleviating the computing pressure.

Finally, we have built a prototype for achieving handover in mobile satellite network. This prototype mainly consists of modified UERANSIM and Open5GS and is driven by real LEO satellite traces including Starlink and Kuiper \cite{Kuiper}. Based on this prototype, we have conducted extensive experiments and results verify that the proposed handover scheme can considerably reduce the handover latency (around 10$\times$) and improve the user-level performance like TCP compared to three existing handover strategies. The detailed implementation of our built prototype for mobile satellite network will be published later.

Additionally, we have also evaluated the performance of the prediction algorithm and the impact of user mobility. The results validate the feasibility of the proposed handover scheme in large-scale mobile satellite network. 

The contributions of this paper can be summarized as:

\begin{itemize}

\item To the best of our knowledge, this work represents the first research efforts to address the high latency problem of handover in mobile satellite network, which is an integral part for 5G and beyond NTN.

\item We for the first time demonstrate how to decouple the interaction with the core network from the handover procedure by leveraging several intrinsic features in LEO satellite network such as predictable satellite trajectory and unique spatial distribution. 


\item We have built an experimental prototype and conduct extensive experiments which is driven by real satellite traces including Starlink and Kuiper. Results verify that the proposed handover scheme can reduce the handover latency by around 10$\times$ compared to three existing handover strategies.  


\end{itemize}
The rest of this paper is structured as follows. Section II introduce the background of the problem and our motivation. Section III gives an overview of our design. Section IV provide detailed explanations of the two aspects of our design. Section V describes our experimental setup and result. Section VI presents a review of related work in the field. Section VII discusses additional considerations and issues related to our work. Finally, Section VIII briefly concludes this work.

\section{Background} \label{chapect 2}
\subsection{LEO Satellite Nework}

In recent years, LEO satellite networks have drawn widespread attentions due to the rapid advance in manufacturing and launch technologies \cite{lin2023fedsn}. Based on the large-scale deployment of LEO satellites, UEs on the ground can enjoy low-latency and high-bandwidth network services with global coverage. Actually, LEO satellite networks can complement and integrate with the traditional terrestrial networks to provide more robust and disaster-resistant communication services. Up to now, it is reported that there have been millions of users using services provided by satellite networks \cite{InvestingStarlink}. 

To achieve real global coverage, ISL have been introduced and testing. In convention, each LEO satellite builds four ISLs with nearby satellites, respectively as two intra-orbit links that connect to the preceding and succeeding satellites in the same orbit and two inter-orbit links that connect to the satellites in adjacent orbits. Based on the ISLs using laser communications, high-bandwidth and stable communication between satellites can be easily achieved.




\subsection{Handover in Mobile Satellite Network} \label{handover in mobile}

\begin{figure}[t!]
	\centering
	\includegraphics[width=0.9\linewidth]{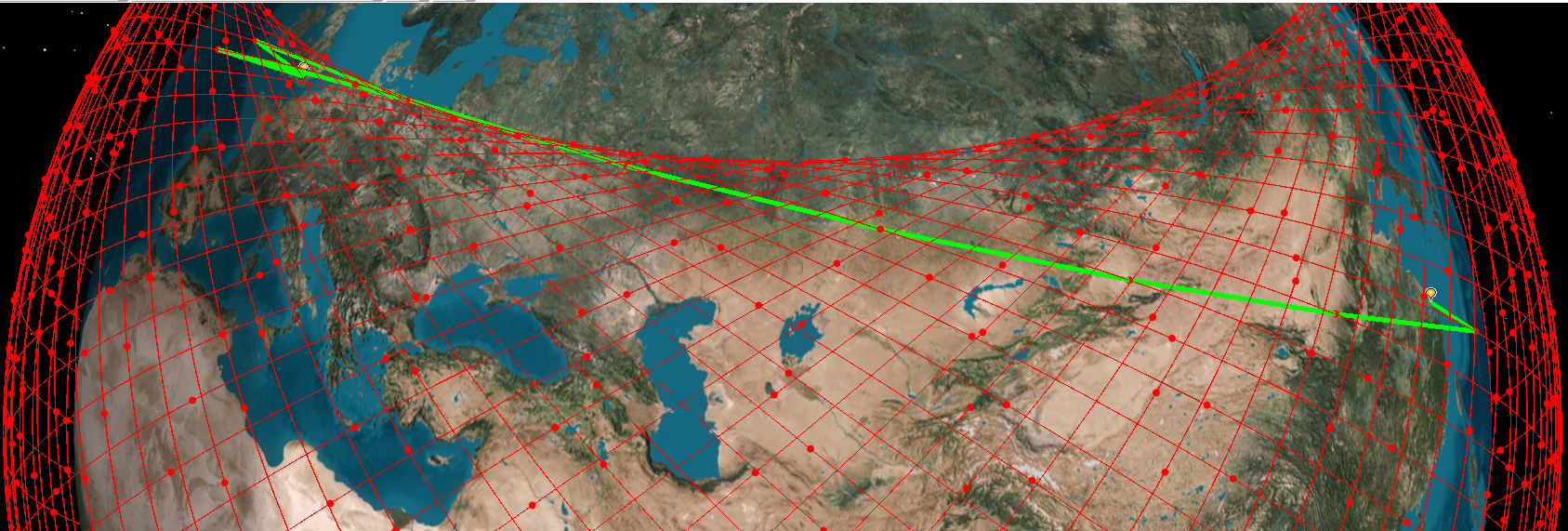}
	\caption{Illustration of handover signaling transmission, e.g., London to Shanghai.}
 \vspace{-0.6cm}
	\label{lts}
\end{figure}

Mobile satellite network follows the handover procedure specified in 5G standard \cite{38821}. There exist two kinds of handover schemes, respectively as Xn-based and N2-based schemes \cite{23501}. By leveraging ISLs, Xn-based handover strategy can provide lower handover latency compared to the N2-based one. High-level Xn-based handover procedure can be summarized as follows. First, UE disconnects the built Radio Resource Control (RRC) link with the source S-gNB and establishes a new RRC satellite-ground connection with the target S-gNB. Secondly, the target S-gNB delivers the handover results to the User Plane Function (UPF) through the ISLs and satellite-ground link, which is a critical part of the core network.   


According to the above framework, there exists a key challenge (i.e., the handover problem) in mobile satellite network. Typically, handover in mobile satellite network differs from the terrestrial network in two main aspects\textemdash which will be illustrated as below. 

\begin{table}[t!] 
	\centering  
	\caption{Orbit information for commercial constellations}  
	\label{Orbit}  
	\begin{tabular}{|c|c|c|c|c|}  
		\hline  
		& & & &  \\[-6pt]  
		&\makecell{Altitude \\ ($km$)}& \makecell{Min. \\ Elevation  ($^\circ$)}&\makecell{Speed \\ ($km/s$)}&\makecell{Avg. Coverage \\ Time ($s$)} \\  
		\hline
		& & & & \\[-6pt]  
		Starlink&550&40&7.8&132 \\
            \hline
		& & & &\\[-6pt]  
		Kuiper&630&35&7.5&187 \\
		\hline
	\end{tabular} \label{Oi}

\end{table}

First, ground UEs experience frequent handover due to the fast travelling speed and small coverage of LEO satellites (compared with GEO satellites). In average, there exists a handover happening every 3 minutes. Actually, the handover interval may be even shorter because of the weather and UE’s access strategy. Table \ref{Oi} depicts the orbit information for two well-known commercial constellations. From this table, we can see that LEO satellites operating at different altitudes have different speeds and thus different time duration for connection. For both of these two constellations, the handover interval is less than 3 minutes because of the limited coverage time for each satellite.


Secondly, another distinguishing feature from the terrestrial network is that the mobile satellite network can provide seamless coverage all over the world even in deserts or oceans, which implies that the distance between UEs and the core network can be much longer than the terrestrial network \cite{6Gwhite,Sateliot}. Consequently, ISLs should be utilized to achieve handover signalling delivery between the access and core networks, which brings much higher latency. For example, Fig. \ref{lts} describes the signaling delivery between Shanghai and London, the distance of which reaches up to 9,000 km. Obviously, this incurs a rather long transmission latency, as well as a large handover latency. Moreover, due to the limited number of ISLs that can be established for each satellite, circuitous transmission path between the access and core networks are common with more ISLs involved, thereby prolonging the propagation and handover latency.


\section{Design Overview}
To mitigate the handover problem in mobile satellite network, we propose a novel scheme for reducing the handover latency by making full use of the predictable satellite trajectory and spatial distribution of satellites. 

\begin{figure}[t!]
    \vspace{-0.6cm}
\centering
   \subfloat[Starlink]{\includegraphics[width=0.45\linewidth]{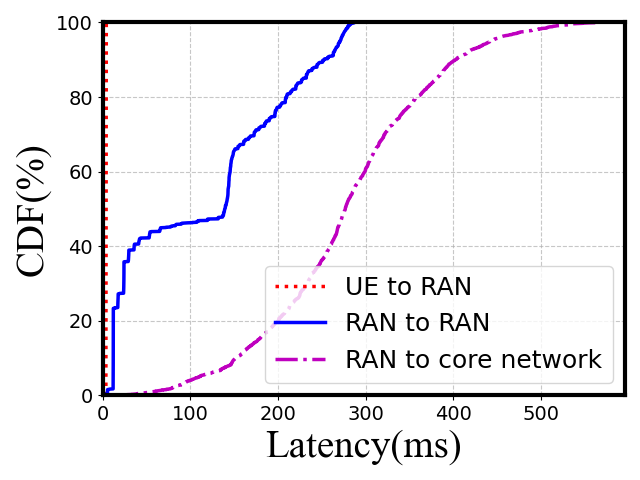}\label{static}}
    \subfloat[Kuiper]{\includegraphics[width=0.45\linewidth]{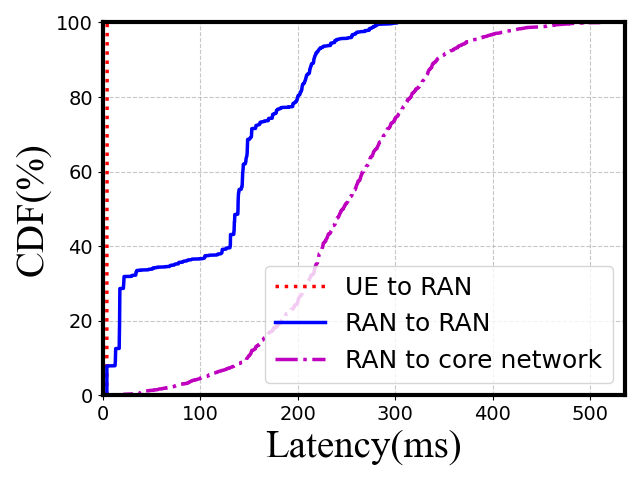}\label{static_Kuiper}}
    \caption{
  Transmission latency of different parts in handover.}\label{SSTAIC}
  \vspace{-0.6cm}
\end{figure}

We divide the overall handover procedure in mobile satellite network into three parts: UE to the Radio Access Network (RAN), RAN to RAN, and RAN to the core network. We conducted preliminary experiments in  to investigate the latency brought by these different steps. As shown in Fig.~\ref{SSTAIC}, we find that the time overhead consumed by the transmission from RAN to the core network dominates the overall handover latency. The reason is that this process consists of passing through multiple ISLs and the satellite-ground link. Therefore, the main aim of this work is to mitigate or even eliminate the interaction between RAN and the core network, thus considerably reducing the handover delay in mobile satellite network.


In this paper, we have redesigned the handover signaling procedure to avoid interactions between the core network and RAN. However, this design introduces two new challenges. The first challenge is related to the synchronization problem between the core network and RAN since there exists no control signalling interaction between them, yet in the standard handover procedure of mobile satellite network there exists much interaction between the core network and RAN to ensure synchronization. The second challenge is the heavy computation overhead on the core network which may paralyze the core network since much trajectory prediction should be performed at the core network side.   


To deal with the first challenge, we propose a fine-grained synchronized algorithm. Specifically, according to predictable trajectory of satellites and weather information, we predict the UE's access satellites at two time points with a fixed interval without interaction with RAN, which then are utilized to determine whether the handover happens. For instance, we employ the simple yet effective binary search to achieve accurate prediction of the handover triggering time. 

We address the second challenge by leveraging the satellite access strategy and the unique spatial distribution of LEO satellites to significantly reduce the number of UEs and satellites required for prediction and hence relieve the computational pressure for the core network.   

Furthermore, we have introduced several extra designs to reduce the handover latency and improve the system robustness. For example, we add a simple constraint (i.e., selecting the access satellite in similar travelling direction with the previous connected satellite) in the satellite access scheme. Later analysis and experiments will demonstrate this easily achievable aim can considerably reduce the handover latency. Simultaneously, we investigate the impact of inaccurate prediction on the performance of the proposed handover scheme. To address its two main causes\textemdash user mobility and deviation in satellite trajectory prediction, we discuss and design corresponding strategies to deal with potential failures.


\section{Handover Design}
In this section, we provide a detailed explanation of our proposed handover scheme. First, we present a comprehensive overview of the proposed handover signaling procedure (\ref{detail_procedure}). 
Then we delve into the two key designs: the synchronized algorithm between RAN and UPF (\ref{Synchronized}) and the computation overhead mitigation for UPF (\ref{satellite_predict}). After that we further optimize the access satellite selection scheme to reduce the handover latency (\ref{Access_selection}). Finally, we introduce how we deal with the inaccurate prediction caused by user mobility and deviations in satellite trajectory prediction. (\ref{users_movement_fail}).
 
\subsection{Detailed Handover Procedure} \label{detail_procedure}

\begin{figure}[t!]
	\centering
	\includegraphics[width=0.92\linewidth]{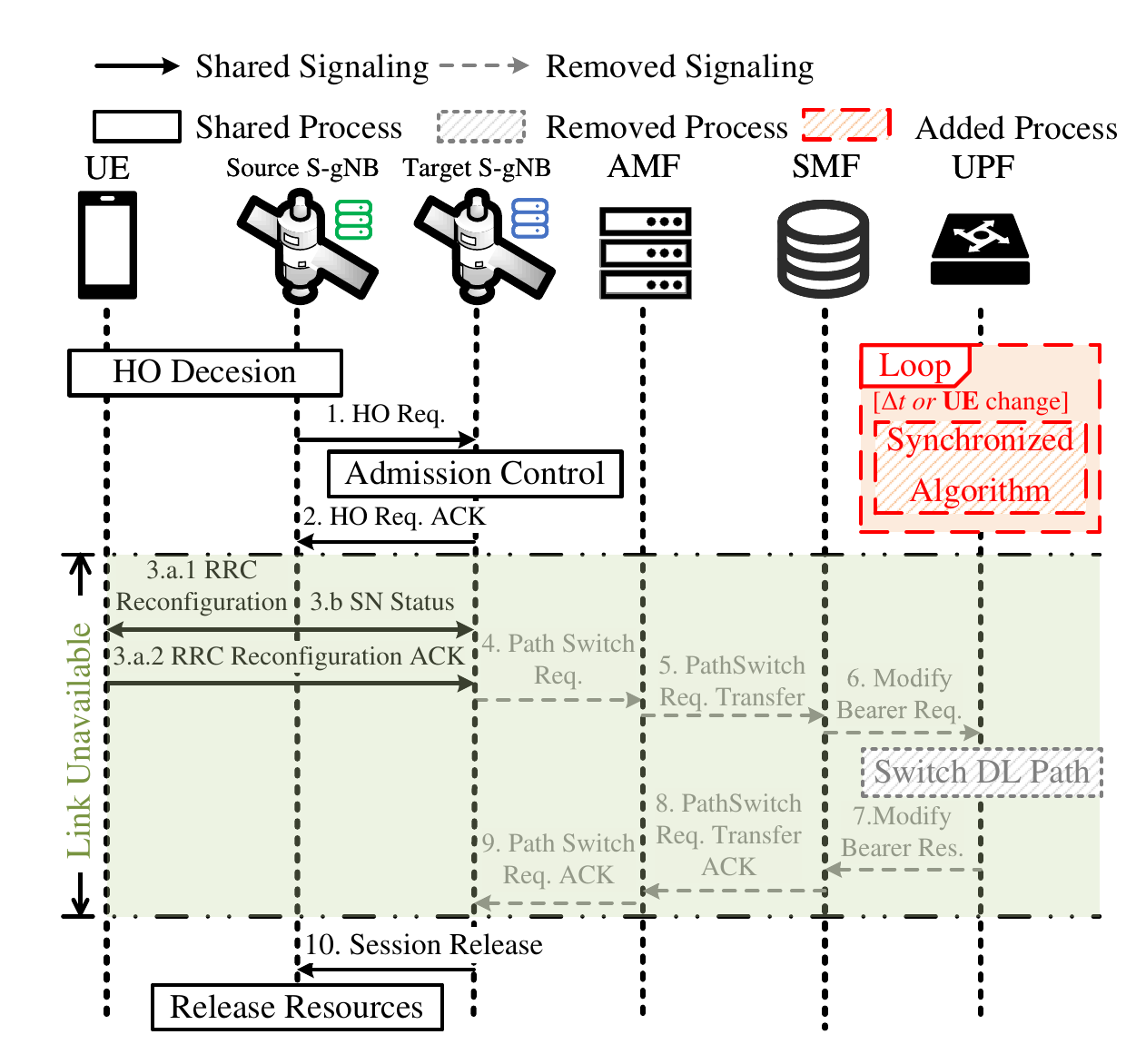}
	\caption{Comparison of standard and our proposed Xn-based handover procedure in mobile satellite network.}
 
	\label{handover_procedure}
 \vspace{-0.6cm}
\end{figure}

As shown in Fig.~\ref{handover_procedure}, similar with the conventional handover procedure, the proposed handover process begins with the handover decision. Handover is triggered by the source S-gNB, which also selects the target S-gNB that be switched to. The following handover procedure can be summarized as below. \\
\textbf{Step 1}. The source S-gNB informs the target S-gNB of the handover decision; \\
\textbf{Step 2}. The target S-gNB prepares for the handover such as pre-allocating channel resources. Then, the target S-gNB confirms the handover with the source S-gNB. After the source S-gNB receives the confirmation, the handover preparation is completed. It should be noted that the link between the core network and UE becomes disconnected from this point. \\
\textbf{Step 3}. The third step can be further divided into two simultaneous operations. One is that the source S-gNB informs the UE of the handover decision and then UE 
disconnects with the source S-gNB and builds a novel RRC connection with the target S-gNB, corresponding to Step 3.a.1 and 3.a.2 in Fig.~\ref{handover_procedure}. The other one is that the source S-gNB transmits the relevant synchronized information with the target S-gNB, including the data to be transmitted and sequence number (SN) (Step 3.b). \\
\textbf{Step 4-9}. The difference between our proposed handover scheme and the standard handover procedure is that the steps 4 to 9 (i.e., control signaling delivery between RAN and the core network) are avoided in our proposed handover scheme, which incurs a large delay latency in mobile satellite network. Instead, we propose a synchronized algorithm at the core network side which can synchronize with RAN even without interacting with RAN.\\
\textbf{Step 10}. At last, the source S-gNB is notified to release the resources. Once this is completed, the handover process is accomplished.

From the above description, the key innovation behind our proposed handover scheme is the synchronized algorithm between RAN and the core network which leverages the predictable trajectory of LEO satellites. However, it is non-trivial to achieve this synchronization yet without control signaling interaction between RAN and the core network. For example, limited prediction frequency may lead to unacceptable delay in handover. Meanwhile, the prediction operation for large quantity of ground UEs poses a huge computating pressure on the UPF. 





In the following subsections, we will provide a detailed description of the synchronized algorithm, as well as how to mitigate the computation overhead, by leveraging two inherent features in LEO satellite networks\textemdash predictable trajectory and unique spatial distribution.

\subsection{Synchronized Algorithm between RAN and  Core Network} \label{Synchronized}

The most fundamental issue of achieving synchronization without control signaling interaction is the prediction of UEs' access satellites at the core network side, whose performance is affected by two factors: relative position and weather \cite{38811,4062836,Satellite_handover}, which can be obtained at the core network. The former can be acquired based on the UEs' location and predictable trajectory of the satellites, while the latter can be obtained from Internet. By leveraging these information and UEs' access strategies, the UEs' access satellites can be accurately determined at the core network.



After obtaining UEs' access satellites, the next issue is the asynchronization problem caused by the coarse-grained prediction. Predicting the users' access satellites at one time point and at a fixed time interval $\Delta t $ is a straightforward approach. However, this mechanism can not bypass the asynchronization problem between the RAN and core network, resulting in additional latency during handover. Since $\Delta t$ is typically in the order of hundreds of milliseconds due to the computing complexity of prediction, the additional latency can severely impede the entire handover process.


Alternatively, we investigate the case of predicting the users' access satellites at two time points simultaneously. We find that this mechanism can solve the above asynchronization problem. Detailed description is shown as follow. Assume that $t_0$ and $t_1 = t_0 + \Delta t$ are two consecutive time points at which the UPF performs predictions. We use $U$ to refer to all UEs served by the UPF, and use $\mathcal{A}_t$ to represent the set of access satellites for all users $U$ at time $t$. By reasonably selecting $\Delta t$, we ensure that at most one handover is triggered for each UE $u \in U$ between $t_0$ and $t_1$. As a result, by comparing $\mathcal{A}_{t_0}[u]$ and $\mathcal{A}_{t_1}[u]$\textemdash the access satellites of user $u$ at time $t_0$ and $t_1$, we can determine whether a handover will be triggered during $t_0$ to $t_1$. If a handover will be triggered, the UPF needs to predict the accurate handover triggering time to minimize the asynchronization duration between RAN and the core network. To avoid any additional delay, the predicted handover triggering time \replaced{$t_p$}{$t_e$} should lie between the handover trigger and the handover completion in the RAN. To achieve this aim, we can employ the simple yet effective binary search method to pinpoint this time point.

It should be noted that the above results are only discussed in common cases. In some special cases when users register, deregister, or travel fast, the corresponding access satellites change accordingly, which should be taken into account in the synchronized algorithm design. 

Consequently, we propose a synchronized algorithm to ensure the synchronization between the RAN and the core network for UPF, as shown in Alg. \ref{Algorithm Synchronized}. Specifically, we define $T$ as the last time point when the periodic update is completed, and $\mathcal{T}_p$ as the prediction set of handover triggering time for each UE. The computation results including $\mathcal{A}_{T}$, $\mathcal{A}_{T+\Delta t}$ and $\mathcal{T}_p$ are stored in the table $\mathcal{R}$. In the subsequent discussion, we will refer to $\mathcal{A}_{T}[u]$ as the \textbf{access satellite} of $u$, and $\mathcal{A}_{T+\Delta t}[u]$ as the \textbf{next-access satellite} of $u$. The proposed synchronized algorithm considers two update cases according to the reason which incurs the update.

\begin{algorithm}[t]
\caption{Synchronized algorithm for the UPF\label{Algorithm Synchronized}}   
{{\small
    \begin{algorithmic}[1]
    \State Initialize $T$, $\mathcal{R}$ 
    \While{True}
    \If {current time $>T$} \Call{Periodic Update}{$T+\Delta t$}
    \EndIf
    \If{localization of $u$ change} \Call{Update UE} {$u,T$} 
    \EndIf
    \EndWhile
    \Procedure{Periodic Update} {$t$} 
    \State Get $\mathcal{A}_{t}$ from $\mathcal{R}$
    \State Based on $\mathcal{A}_{t}$, predict $\mathcal{A}_{t+\Delta t}$ according 
    \State According to $\mathcal{A}_{t}$ and $\mathcal{A}_{t+\Delta t}$, get $\mathcal{T}_p$ with binary search 
    \State Wait until $t$, update $\mathcal{R}$ 
    \State $T = t$
    \EndProcedure
    
    \Procedure{Update UE} {$u,t$} 
    \State Calculate \replaced{$\mathcal{A}_{t}[u]$}{$\mathcal{A}_{t+\Delta t}[u]$}, $\mathcal{A}_{t+\Delta t}[u]$, and $\mathcal{T}_p[u]$
    \State Update $\mathcal{R}$ 
    \EndProcedure
    
    \end{algorithmic}
    }
    }
\end{algorithm}

\begin{itemize}
\item \textbf{Periodic update caused by satellite movement}: Due to the fast travelling speed of LEO satellites, the synchronized algorithm iteratively performs a periodic update with a periodicity of $\Delta t$. Below, we provide a step-by-step description of the periodic update for $T+\Delta t$ as an example. First, we obtain the access satellites of all UEs at $T+\Delta t$\textemdash $\mathcal{A}_{T+\Delta t}$ from $\mathcal{R}$, and then predict the access satellites at $T+2\Delta t$\textemdash $\mathcal{A}_{T+2\Delta t}$ according to the predicted trajectory of LEO satellites. According to the difference between $\mathcal{A}_{T+\Delta t}$ and $\mathcal{A}_{T+2\Delta t}$, we employ binary search algorithm to calculate the set of predicted handover triggering time $\mathcal{T}_p$. \added{Specifically, for each UE $u$ with distinct values of $\mathcal{A}_{T+\Delta t}[u]$ and $\mathcal{A}_{T+2\Delta t}[u]$, we compute their associated access satellites at the intermediate time instant, i.e., $T+1.5 \Delta t$, to halve the prediction error for access time. This process is repeated iteratively until the error is less than the time required for the handover procedure at RAN.} To avoid the premature update, we update the table $\mathcal{R}$ when the time reaches $T+\Delta t$.


\item \textbf{Update caused by UE}:
When the UE $u$ undergoes a localization change, which may be caused by the registration, deregistration, or movement, the algorithm will also perform an update for UE $u$. At this point, our proposed synchronized algorithm updates $\mathcal{A}_T$, $\mathcal{A}_{T+\Delta t}$ and $\mathcal{T}_p$ in the table $\mathcal{R}$.

\end{itemize}

Next, we describe how we put the synchronized algorithm into the mobile satellite network system to achieve synchronization between RAN and the core newtork.

\begin{figure}[t]
        
	\centering
	\includegraphics[width=3.5in]{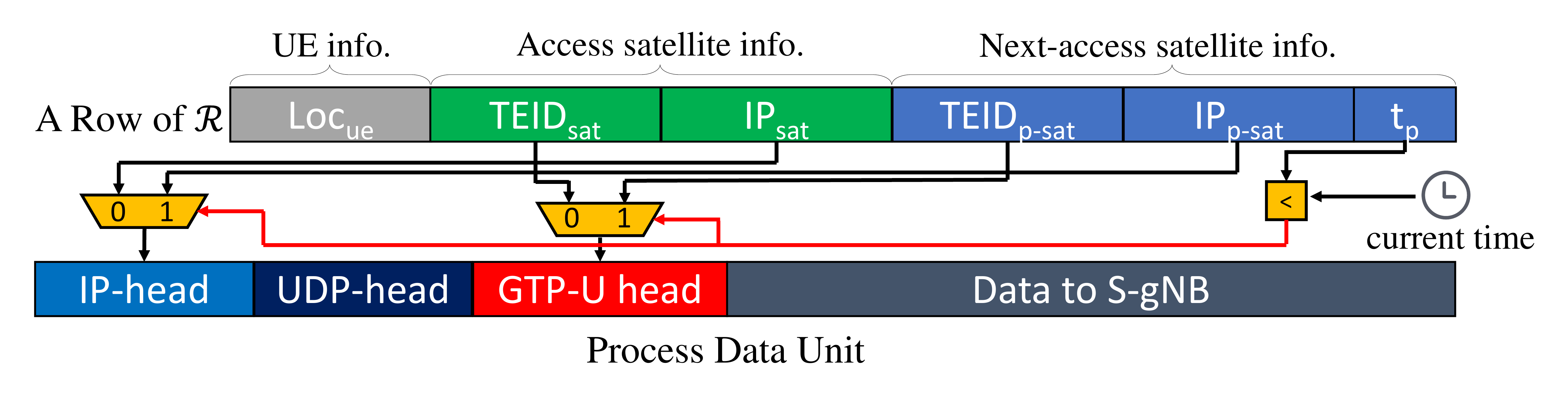}
	\caption{Structure and Operation of UE-satellite relationship.}
	\label{relationship}
 \vspace{-0.6cm}
\end{figure}

Actually, main modifications take place in UPF. As mentioned above, UPF maintains the table $\mathcal{R}$ and updates it through the synchronized algorithm. As shown in Fig. \ref{relationship}, $\mathcal{R}$ consists of user information corresponding to $U$, access information corresponding to $\mathcal{A}_{T}$, and next-access information corresponding to \replaced{$\mathcal{A}_{T+\Delta t}$}{$A_{T+\Delta t}$} and $\mathcal{T}_p$. Notably, if there is no handover for user $u$ between $T$ and $T+\Delta t$, the corresponding next-access satellite remains the same as the access satellite, and $t_p = \added{\mathcal{T}_p[u] =} 0$. When there is a packet that needs to be transmitted to UE, UPF can obtain related information such as $TEID_{sat}$ from table $\mathcal{R}$ and then feed them into the protocol header. In addition, in the UE registration/deregistration procedure, UPF triggers the synchronized algorithm to update $\mathcal{R}$.

To maintain the synchronization, a few modifications are made at S-gNB. In particular, the UEs' location information is required to be fed into the GTP Extension Header of uplink data, in order to notify the UPF of changes in UEs' location. 

Finally, we explore the appropriate value set for the update interval $\Delta t$,  which should ensure that two consecutive handovers for any stationary UE do not occur within $\Delta t$. A feasible solution is to set $\Delta t$ to be less than constellation's minimum service time, which is influenced both by the satellite constellation configuration and the perhaps additional limitation when selecting access satellites (to avoid excessive frequent handovers). Based on the analysis of existing satellite constellation configurations and actual measurements \cite{neinavaie2022unveiling,starlink,Kuiper,oneweb}, we set the update interval $\Delta t$ to 5 seconds. \added{Consequently, by iterating the binary search process 9 times, we can attain a prediction accuracy as precise as 10 milliseconds, meeting the specified requirements.}

\subsection{Optimization for Prediction Algorithm} \label{satellite_predict}

According to the previous discussions, each UPF needs to predict the access satellites for all UEs it serves every $\Delta t$. A directly prediction approach, which considers all satellites in the constellation for every UE to predict the access satellite, would impose significant computational pressure and make it difficult to complete within the specified $\Delta t$.

To address this challenge, we introduce the fast access satellite prediction algorithm. First, we take into account the access strategies of UEs to reduce the number of users that need to be considered. We categorize the existing access strategies into two types: \textbf{consistent} and \textbf{flexible} \cite{4062836,Satellite_handover}. The former maintains a connection until the user leaves the service coverage area, while the latter may handover even when UE is within its access satellite's coverage area. For UEs with consistent access strategies, the algorithm simply checks if the previously connected satellite can still provide service, significantly reducing the number of UEs required consideration. 

Additionally, we reduce the number of satellites in each UE's access satellite prediction, according to satellites' spatial distribution. We divide the Earth into several continuous rectangular blocks based on the satellite's service radius of the satellites, and assign satellites into corresponding blocks according to their position. As a result, for each UE, only nearby satellites required computation, which greatly reduces computation complexity.


\begin{algorithm}[t]
\caption{Fast access satellite prediction algorithm\label{Algorithm 1}}   
{{\small
    \begin{algorithmic}[1]
    \Require$\mathcal{A}_t$, $U$
    \Ensure $\mathcal{A}_{t'}$
    \State Predict $S_{t'}$, the satellites' position at time $t'$, 
    \State Initialize UE candidate $U_C$, $A_{t'}$
    \ForAll{$u_i \in U$}
    \If{Access strategy is consistent} 
        \If{satellite $s=\mathcal{A}_t(u_i)$ is not available for $u_i$}
            \State  $U_C = U_C \cup \{u_i\}$
        \Else
            \State $\mathcal{A}_{t'}(u_i) = \mathcal{A}_t(u_i)$
        \EndIf
    \ElsIf{Access strategy is flexible}
        \State  $U_C = U_C \cup \{u_i\}$
    \EndIf
    \EndFor
    \State Create geographical blocks $B=\{B_1,B_2,...,B_n\}$
    \State Initialize satellite sets $S^i$, $i = 1,2, ..., n$
    \ForAll{$s_j \in S_{t'}$}
    \State Identifying the block $B_i$ where $s_j$ is located
    \State $S^i = S^i \cup \{s_j\}$
    \EndFor
    \ForAll{$u_j \in U_C$}
    \State Identifying the block $B_i$ where $u_j$ is located
    \State From $S^i$ and its neighbour, find the access satellite $s$ of $u_j$ 
    \State $\mathcal{A}_{t'}(u_i) = s$ 
    \EndFor
    \State \Return $\mathcal{A}_{t'}$
    \end{algorithmic}
    }

    }

\end{algorithm}
The pseudo code of the proposed algorithm is shown in Alg. \ref{Algorithm 1}. Specifically, the algorithm can be divided into 4 steps as following: \\
\textbf{Step 1}: Based on the ephemeris, we predict $S_{t'}$, the satellites' position at time $t'$. \\
\textbf{Step 2}: To different selection strategies, we employ distinct approaches to construct the UE candidate set $U_C$, which consists of UEs that may undergo handover. For UEs with consistent strategy, we only add those whose access satellite is not available at $t'$ into the set $U_C$ for further processing. For UEs with flexible strategy, we add all UEs to the set $U_C$. \\ 
\textbf{Step 3}: We divide the Earth into several rectangular blocks with the satellite service radius as side length. All satellites are allocated to correspondingly blocks according to their position at time $t'$. \\
\textbf{Step 4}: For each user in $U_C$, we calculate the access satellite it should be connected to at time $t'$ based on its access strategy.

\subsection{Optimization on Access Satellite Selection} \label{Access_selection}
To further reduce the overall handover latency, we focus on minimizing the latency between source S-gNB and target S-gNB, which is the time taken for the handover process. Based on the preliminary experiments shown in Fig. 3, we can observe a sudden increase in latency in 30-40 ms. The main reason is that the satellite switches in contrary directions (i.e., handover from a northern-direction satellite to a southern-direction satellite, or vice versa) \cite{chen2021analysis,zhang2022enabling}.
To this end, we make an additional constraint that handovers can happen only between satellites in the similar-direction ones, which is expected to reduce the propagation delay between satellites by at most more than 200ms.

\subsection{Inaccurate Prediction} \label{users_movement_fail}
\begin{figure}[t!]
	\centering
	\includegraphics[width=\linewidth]{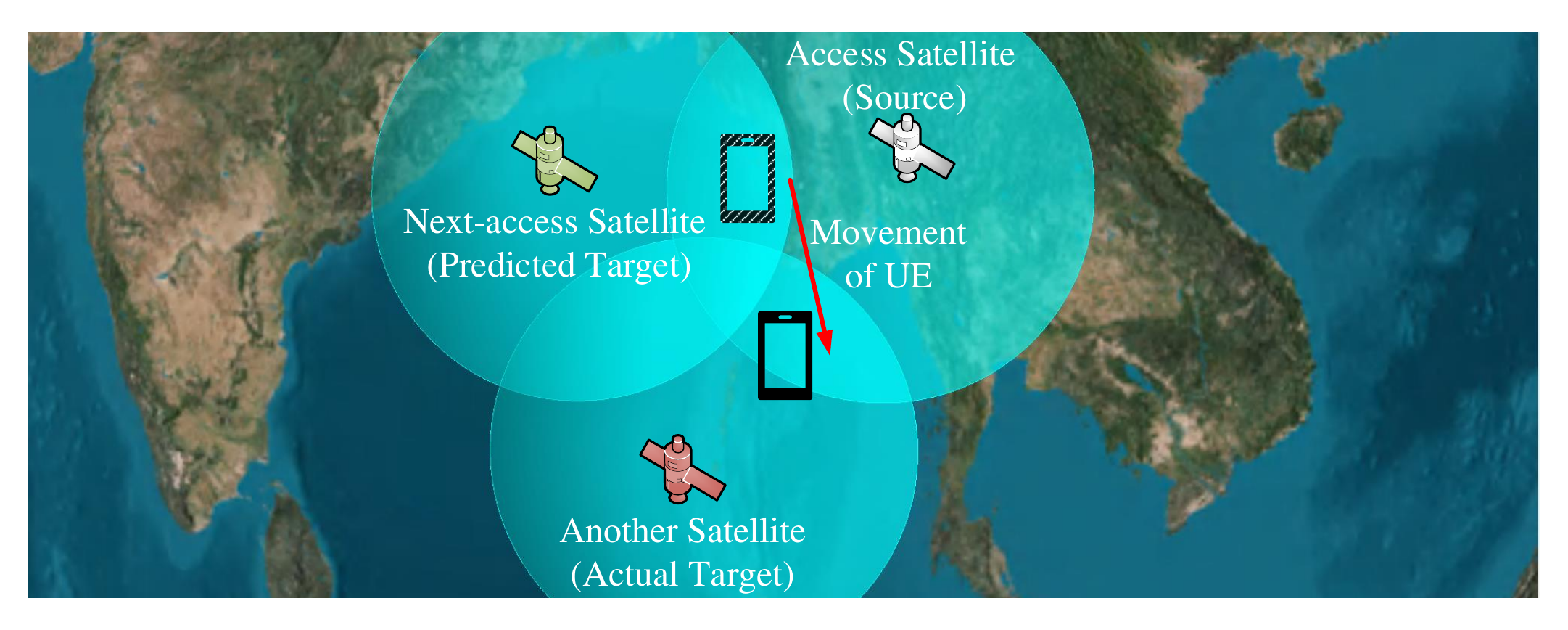}
	\caption{Illustration of abnormal handover caused by user mobility.}
	\label{uemobile}
 \vspace{-0.6cm}
\end{figure}

User mobility and the deviations in satellite trajectory prediction may cause inaccurate predictions of UEs' access satellites, resulting in what we refer to as ‘abnormal handovers’, where the switching target is not the predicted satellite. This will negatively affect the availability of our proposed scheme. Therefore, we discuss the impact of these two causes and propose solutions. 

On one hand, in the context of users, despite their significantly lower speed compared to LEO satellites, there exist some cases where user mobility results in inaccurate prediction at the core network and leads to an abnormal handover. Such cases may occur when user move at high speeds and is located at the overlapping region of the satellites' service coverage, as illustrated in Fig. \ref{uemobile}. \deleted{To deal with such abnormal handover, the S-gNB triggers the standard 5G NTN handover procedure, transmitting control signaling to the core network and ultimately updating the UE information at the UPF. }
It is crucial to highlight that these scenarios are infrequent, even for UEs who are moving at high speeds. Consequently, it can be regarded that user mobility has little impact on the user experience during the handover process. In \ref{results of Experiment}, we will delve into the probability of abnormal handover using experimental data.

\replaced{The perturbations in satellite motion may result in errors of several kilometers in trajectory prediction based on daily-updated ephemeris data \cite{orbitprediction}, consequently causing abnormal handovers with a probability of approximately $10^{-3}$. To tackle this problem, one solution is to employ short-term trajectory prediction based on minute-level updated satellite ephemeris data \cite{rs15010133},  whose prediction precision reaches 10 centimeters.}{On the other hand, the deviations in satellite trajectory prediction will also cause inaccurate prediction. The usual satellite trajectory prediction based on daily-updated ephemeris data may lead to errors of several kilometers, resulting in abnormal handovers with a probability of approximately $10^{-3}$ \cite{orbitprediction}. To tackle this problem, we employ short-term trajectory prediction based on minute-level updated satellite ephemeris data, whose prediction precision reaches 10 centimeters \cite{rs15010133}; thus, avoiding inaccurate predictions. }

\added{To deal with abnormal handovers, the S-gNB triggers the standard 5G NTN handover procedure, transmitting control signaling to the core network and ultimately updating the UE information at the UPF. }
\section{Experiments}

\subsection{Experimental Setup}

\begin{figure}[t!]
\centering
   \subfloat[]{\includegraphics[width=0.22\textwidth]{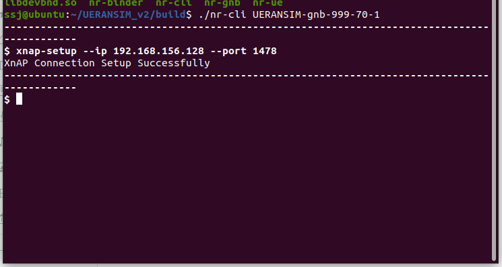}\label{Xnap_Establish_fig}}
   \hspace{0.1in}
    \subfloat[]{\includegraphics[width=0.22\textwidth]{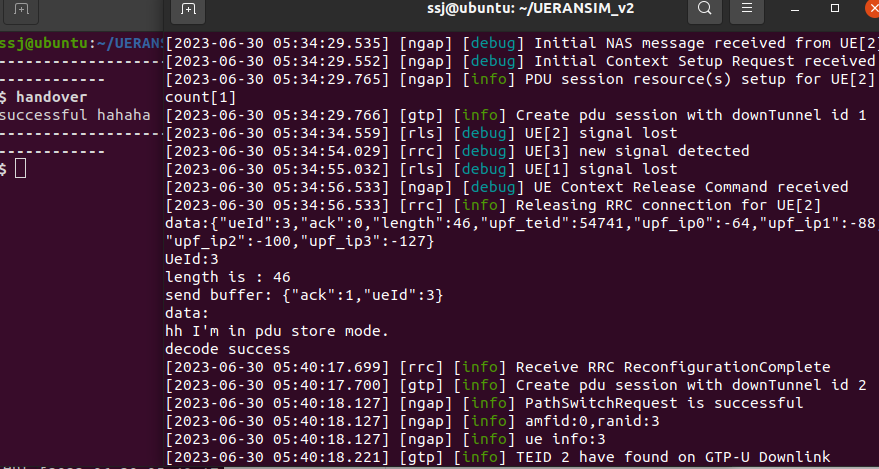}\label{handover_success}}
    \caption{
  Developed Xn-based handover: (a) Xnap connection buildup and (b) Handover success.}\label{Xn-based}
  \vspace{-0.6cm}
\end{figure}

\textbf{Satellite Constellation}: First, experiments are carried out driven by the Starlink and Kuiper constellation traces that are obtained from \cite{celestrak}. And we can also get the information of ground stations from \cite{StarlinkStatus}. 
Finally, we have built a platform for simulating real-trace dynamics of LEO satellite constellation using skyfield \cite{skyfield} based on these constellation and ground stations data.

\textbf{System Prototype}: Driven by the above platform, we have built a prototype by combining UERANSIM \cite{ueransim} with Open5GS \cite{open5gs} for 5G and beyond non-terrestrial network, which achieve handover in mobile satellite network. UERANSIM is a widely utilized simulator for both UE and S-gNBs implementation in 5G network, while Open5GS is employed for implementing the 5G core network. Following the 5G standard signaling flow, we have made modifications to UERANSIM to support Xn-based handover, as shown in Fig. \ref{Xn-based}. The built prototype operates on a commodity laptop with 2.5 GHz CPU core and 16 GB RAM.    


\textbf{Comparison Schemes}: We compare our proposed handover scheme with the following three handover schemes, in order to demonstrate the higher efficiency of the proposed handover strategy.
\begin{itemize}
\item \textbf{NTN} refers to the standard handover process specified in 5G NTN, which is described in \ref{handover in mobile}. We compare the proposed handover scheme with the NTN handover scheme to show the performance improvement brought by the modified handover. 

\item \textbf{NTN-GS} refers to the handover procedure assisted by nearby ground stations, which implies that ground stations close to the LEO satellite are leveraged to record handover information, inspired by the handover strategies designed in IP satellite network \cite{dong2018mianxiang,7811041}. Thus, the controlling signaling for handover does not need to be transmitted to core network and then the handover latency can be reduced.


\item \textbf{NTN-SMN} refers to the handover procedure assisted by nearby space network (SMN) \cite{9755268}. Similar with the NTN-GS handover scheme, satellites act as the access and core network simultaneously in this strategy. In this way, the handover latency can also be reduced.

\end{itemize}

\subsection{Experimental Results} \label{results of Experiment}

\begin{figure*}[t]
\centering
   \subfloat[flexible strategy, similar direction, Starlink]{\includegraphics[width=0.3\textwidth]{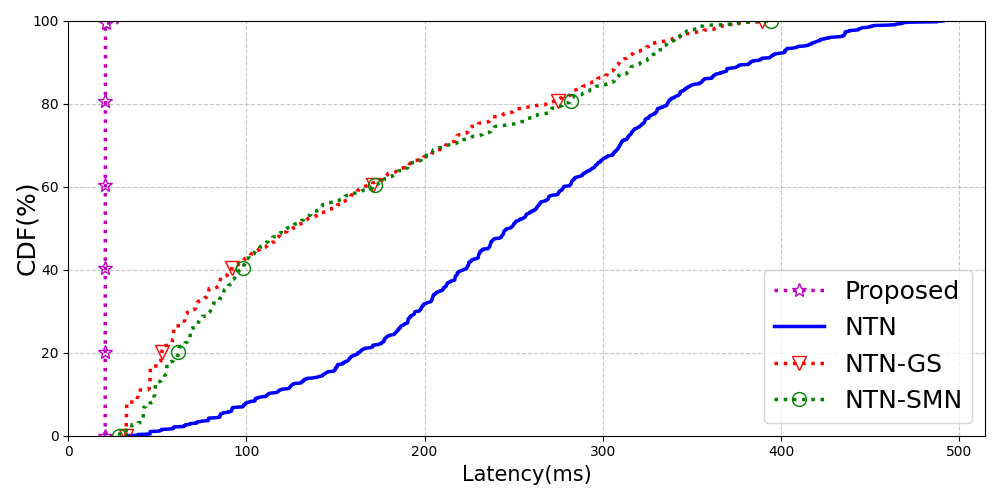}\label{latency}}
    \subfloat[flexible strategy, all direction, Starlink]{\includegraphics[width=0.3\textwidth]{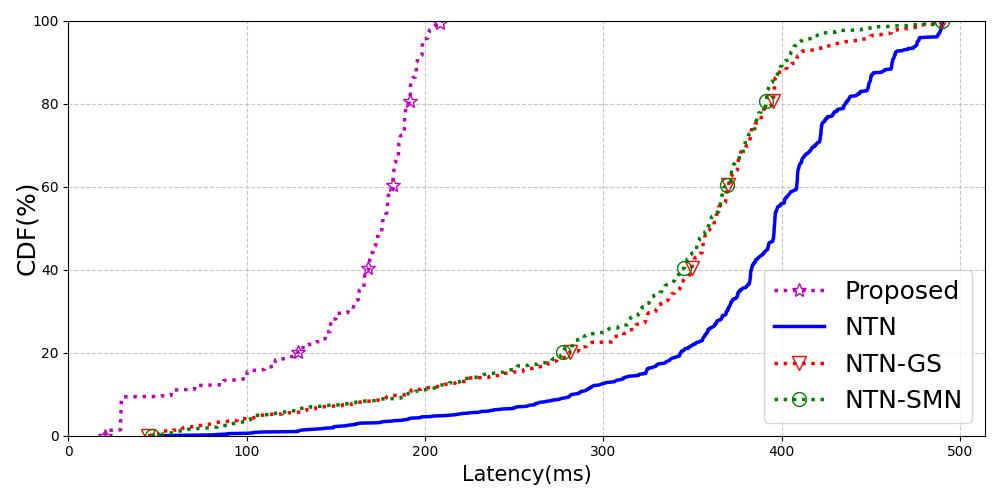}\label{latency_all_direction}}
    \subfloat[consistent strategy, similar direction, Starlink]{\includegraphics[width=0.3\textwidth]{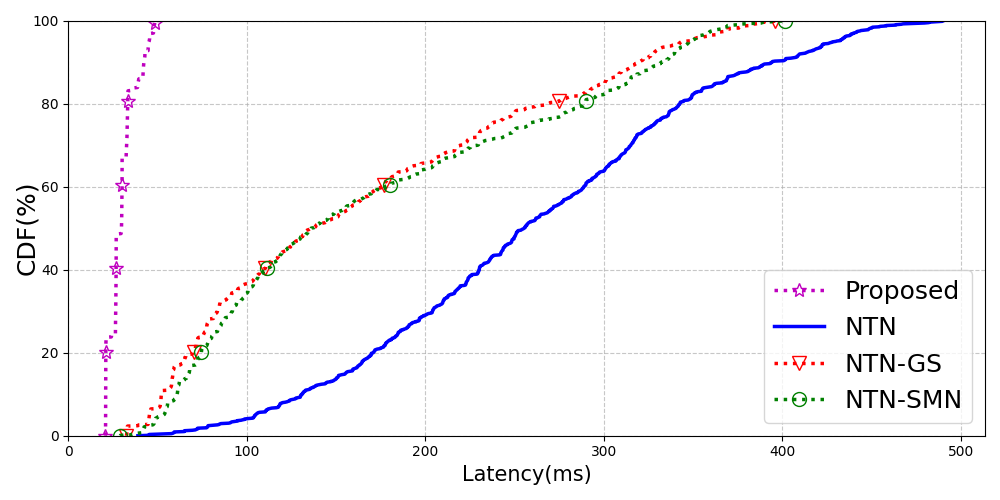}\label{latency_consistent}}
    \quad
    \subfloat[consistent strategy, all direction, Starlink]{\includegraphics[width=0.3\textwidth]{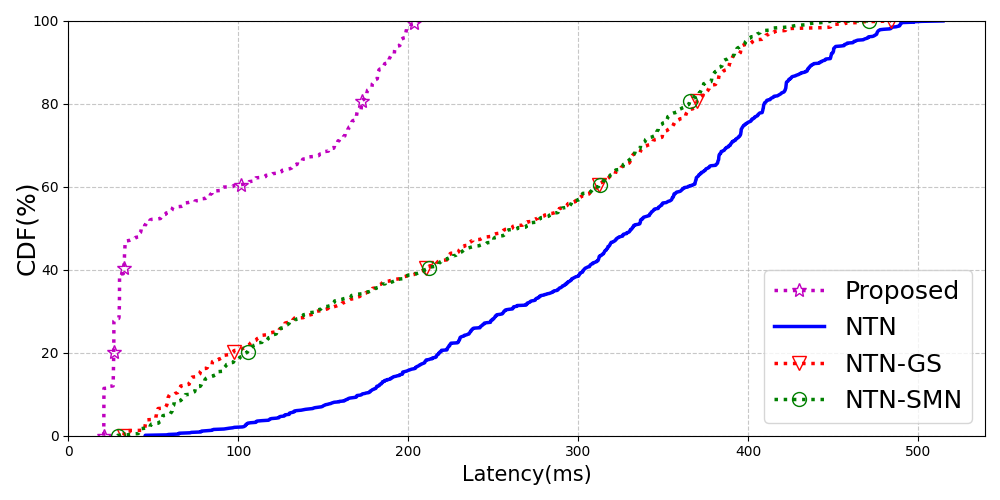}\label{latency_consistent_all_direction}}
    \subfloat[flexible strategy, similar direction, Kuiper]{\includegraphics[width=0.3\textwidth]{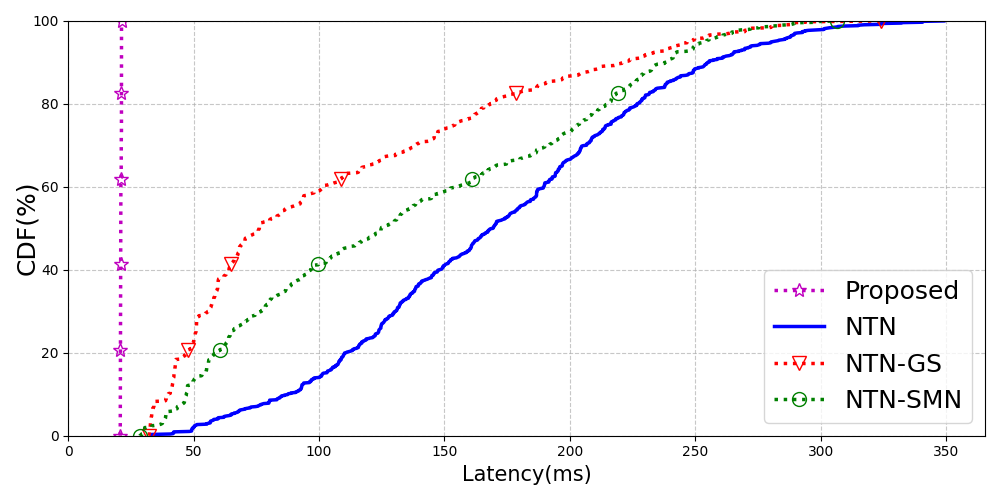}\label{latency_Kuiper}}
    \subfloat[consistent strategy, similar direction, Kuiper]{\includegraphics[width=0.3\textwidth]{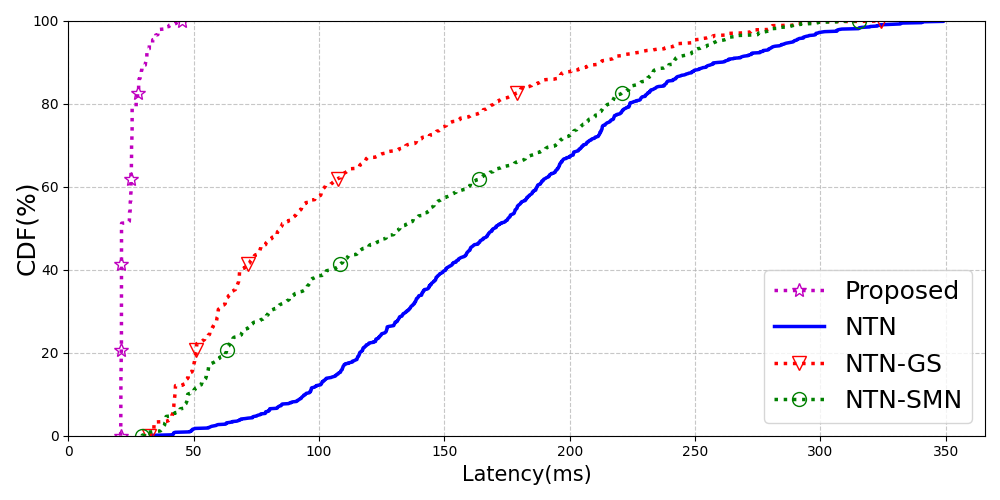}\label{latency_consistent_Kuiper}}
    \caption{Comparison of handover latency using different access satellite selection strategies and constellations.} \label{latency_all} 
    \vspace{-0.5cm}
\end{figure*}
\noindent \textbf{Handover Latency:} As shown in Fig. \ref{latency_all}, we compare the latency of different handover strategies using different access satellite selection strategies and constellations to demonstrate the influence of these metrics. It can be observed that our proposed handover scheme shows a superior performance in terms of latency compared to the three other handover schemes covering all scenarios, which includes both flexible and consistent strategies and two constellations. More specifically, as shown in Fig.~\ref{latency}, the handover latency based on the proposed handover scheme is 20.87 ms on average, which is much shorter than the handover latency of 250 ms based on the NTN strategy. Meanwhile, it is much shorter than the handover latency based on the two other optimized handover schemes (i.e., NTN-GS and NTN-SMN), which is 153 ms and 158.5 ms, respectively.  

Fig.~\ref{latency} and Fig.~\ref{latency_all_direction} describes the handover latency using different access satellite selection strategies. We can observe that the average handover latency increases by around 6.1$\times$ without optimizing the access satellite selection process compared to the proposed handover scheme, which implies that the simple yet effective optimal access satellite selection scheme proposed in the handover procedure can assist in dramatically reducing the handover latency. This performance gain can be attributed to the transmission delay decrease between satellites when satellites are between the same-direction satellite cluster, since inter-satellite information transfer is unavoidable as S-gNBs need to exchange information. Meanwhile, we also observe similar performance based on other handover strategies, highlighting the necessary requirement for the access satellite selection optimization.        



The comparison of Fig.\ref{latency} with Fig.\ref{latency_consistent}, reveals the performance difference of different handover triggering criteria. It can be observed that the handover latency based on the flexible strategy is a little shorter (around 10 ms) than the consistent strategy. The reason is that the source satellite is closer to the target satellite using the flexible strategy as the handover triggering criterion compared to the case of the consistent strategy. However, the flexible strategy brings more handover compared to the the consistent strategy. Consequently, the handover triggering criterion is a valuable research direction which should account for the requirements of applications.  

Finally, we evaluate the performance of the proposed handover scheme under different satellite constellations such as Starlink and Kuiper. 
First, we can observe the proposed handover strategy outperforms the other three handover schemes irrespective of the kind of satellite constellation, as demonstrated in Fig.~\ref{latency} and Fig.~\ref{latency_Kuiper}. This is because the proposed handover scheme achieves handover latency reduction mainly leveraging the predictable travelling trace of LEO satellites and thus irrelevant with the constellation. Secondly, the handover latency exhibits a little difference (about 10\% in average) under Starlink and Kuiper constellations, which can attributed to the difference of constellation configurations such as inter-orbit and intra-orbit distances and satellite altitudes. Thirdly, we can also observe that there exists slight handover latency difference when using different access satellite selection strategies in Kuiper satellite constellation, which is verified by comparing Fig.~\ref{latency_Kuiper} with Fig.~\ref{latency_consistent_Kuiper}.

In conclusion, the proposed handover scheme shows a superior performance over the three other handover schemes in terms of handover latency under various conditions.
\begin{figure}[t!]
    \vspace{-0.3cm}
\centering
   \subfloat[Stalling Time]{\includegraphics[height=0.16\textwidth]{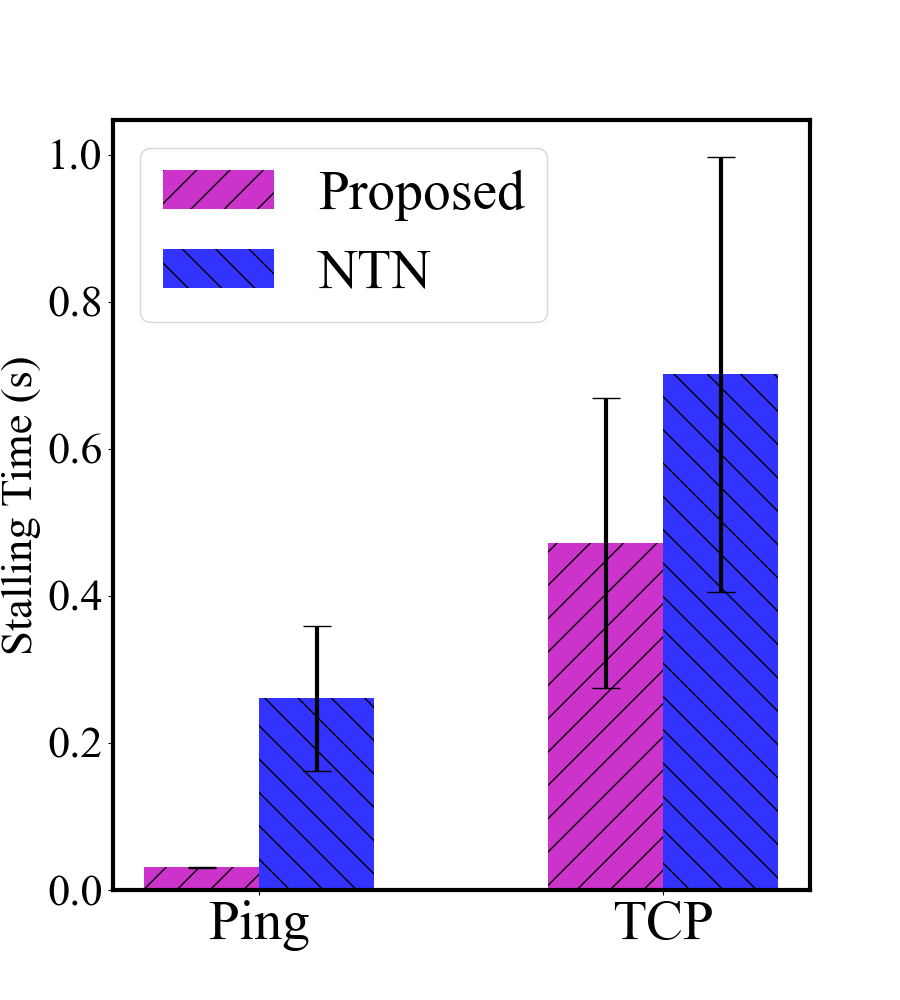}\label{sta}}
    \subfloat[Ping in NTN]{\includegraphics[height=0.16\textwidth]{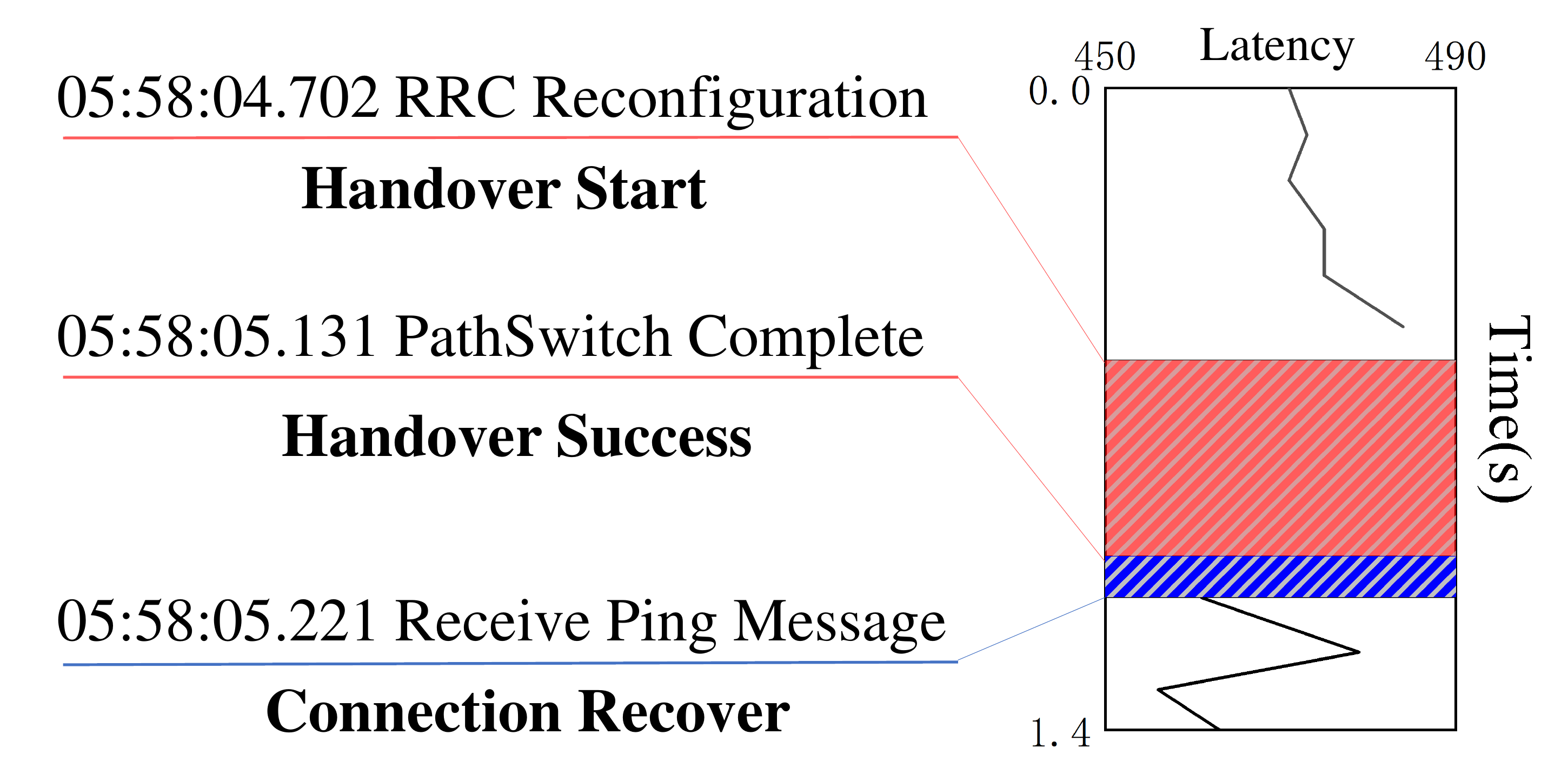}\label{ping detail}}
    \caption{User level performance.}\label{User level}
    \vspace{-0.6cm}
\end{figure}

\noindent \textbf{User-level Performance:} Furthermore, we have compared the user-level performance respectively brought by the proposed handover scheme and 5G NTN handover strategy. As shown in Fig. \ref{User level}, the stalling time based on the proposed handover scheme can be reduced by 89\% compared to 5G NTN handover strategy. Meanwhile, in the case of TCP flows, our proposed handover scheme can decrease the stalling time by 33\% compared to 5G NTN handover, as demonstrated in Fig. \ref{sta}. This is because handover affects the user-level performance, where large handover latency prolongs the delay of user-level applications and then deteriorates the overall performance.   

In the meantime, the reason behind the limited performance improvement in the case of TCP flows is that there exists a three-way handshake mechanism incurring substantial time overhead, which is exacerbated by the long propagation latency between inter-satellite. To illustrate in detail, we delve into details of the ping procedure in mobile satellite network. As shown in Fig.~\ref{ping detail}, the stalling time in the link recovery process can be divided into two steps. First, the user terminal connects with the target S-gNB. Secondly, after a certain duration, the connection between the user and the server is restored. The proposed handover scheme can considerably reduce the time delay of the first step, i.e., the time taken for the link to recover. However, the overhead of the second step is primarily caused by the delay between inter-satellite, which is beyond the scope of this paper and thus resulting in the limited performance improvement in terms of stalling time.                                                                                                                  

\begin{figure}[]
	\centering
	\includegraphics[width=\linewidth]{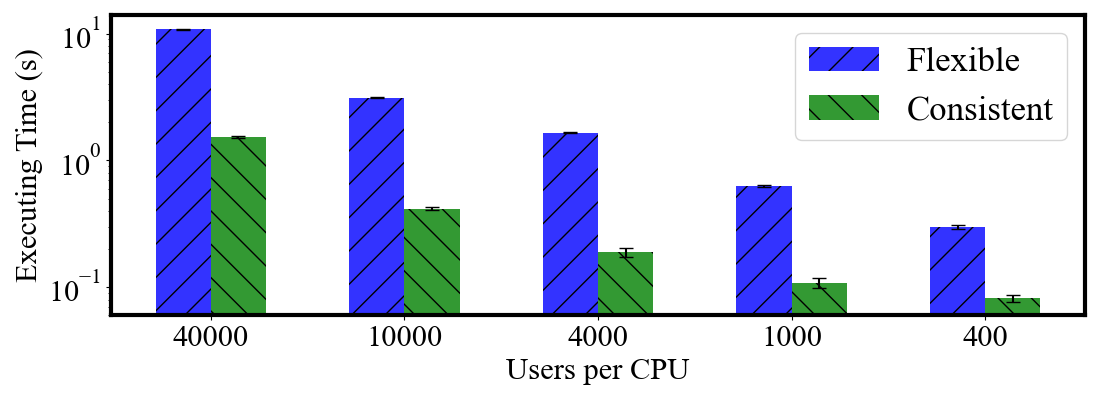}
	\caption{Performance of fast access satellite prediction.}
	\label{alogrithm}
 \vspace{-0.6cm}
\end{figure}

\noindent \textbf{Fast Access Satellite Prediction:} Next, we investigate the performance of the proposed fast access satellite prediction algorithm with different numbers of users. It can be observed from Fig.~\ref{alogrithm} that the time overhead using the consistent access strategy is much lower than the flexible strategy since there exists much less handover in the case of the consistent access strategy than the flexible strategy. Meanwhile, the results that the computing time increases with the number of users are expected. 

According to the deployment of existing commercial mega-constellations, a ground station serves around 10,000 users in average \cite{InvestingStarlink}. When the number of users is 10000, the computing time is around 2 seconds when using a commodity laptop, which is within the time requirements of handover. Even it is predicted that the number of users in satellite networks will increase significantly in the future, thereby intensifying the computing pressure of the ground stations. Actually, this challenge can easily overcome by equipping the ground stations with more high-performance hardware.



\begin{figure}[]
	\centering
	\includegraphics[width=0.9\linewidth]{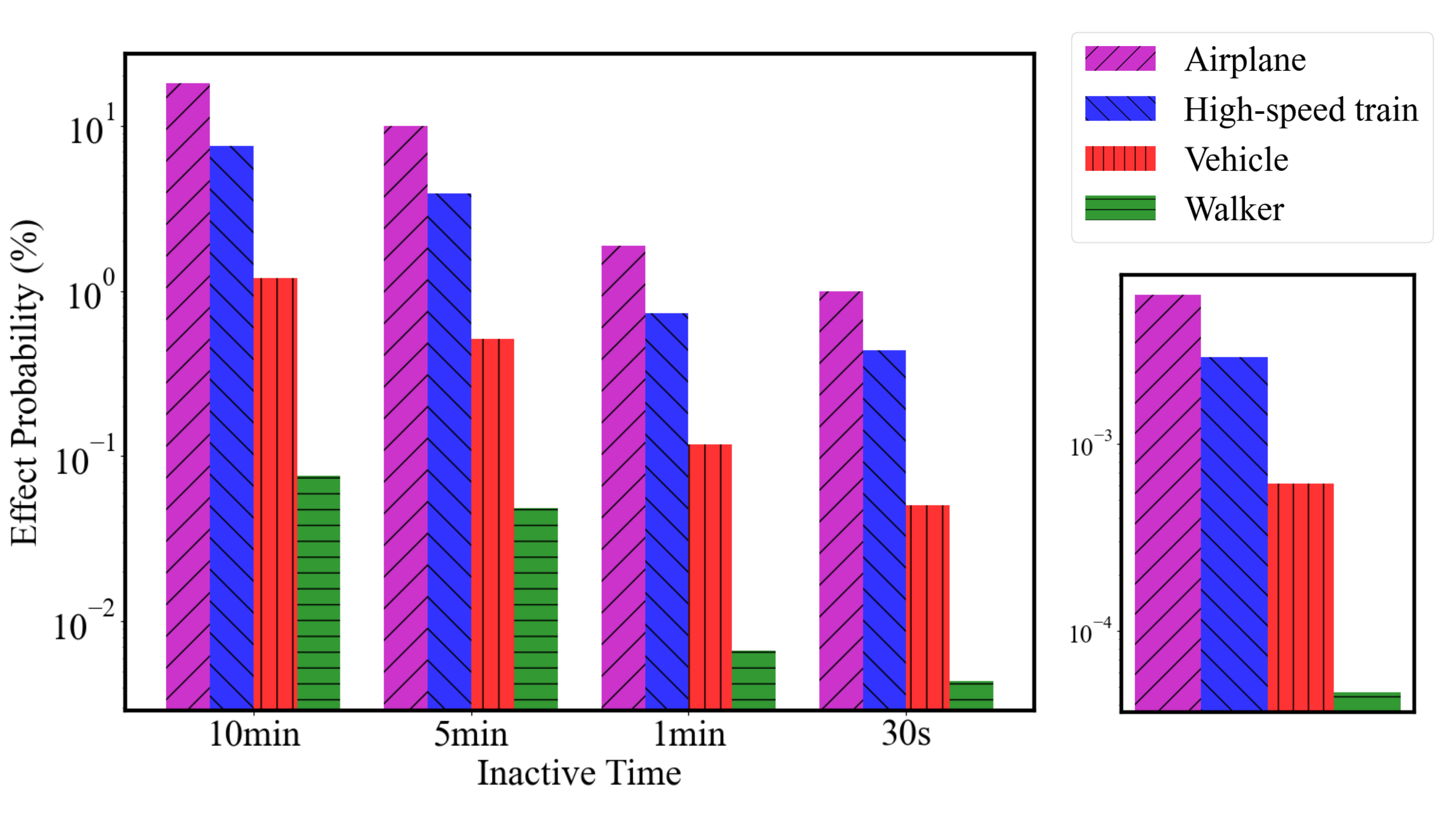}
	\caption{Impact of user mobility. (left: user is inactive; right: user is active.)}\label{fail}
  \vspace{-0.6cm}
\end{figure}


\noindent \textbf{Impact of User Movement:} Finally, we have conducted experiments to evaluate the influence of user movement (e.g., users move at different speeds in both active and inactive scenarios) on handover performance, referred to as ‘abnormal handover’ defined in \ref{users_movement_fail}. As shown in Fig. \ref{fail}, for active ground users, the probability of ‘abnormal handover’ is very low such as approximately $10^{-6}$ for users on high-speed airplanes, and even lower in other scenarios. In the case of walking speed, the abnormal probability is in the order of millionth, which can be considered negligible.


On the other hand, for users who are in high-speed motion and remain inactive for an extended period, the probability of triggering an 'abnormal handover' is relatively higher. For example, after 10 minutes of inactivity, users on airplane have an 18\% probability of experiencing an abnormal handover, while users travelling on high-speed train have a 7\% abnormal probability. Actually, the delay cost of such an abnormal handover is equivalent to the time overhead incurred by the standard 5G NTN handover. Thus, in these cases, the abnormal handover scheme can be regarded as the standard 5G NTN handover. 

\section{Related Work}
The problem of handover in satellite network has been extensively discussed. Considering the satellite network's topology is a natural approach \cite{8926396,9003306}. One mainstream method is the virtual node method \cite{8926396}, in which partition mappings satellites and ground stations into geographic region. However, this method only works in polar orbit constellation networks and is not applicable to mainstream constellation configurations, such as the Walker constellation primarily used by Starlink. Another mainstream method is based on calculating all possible network topology states to predict the network topology at any given moment \cite{9003306}. However, this method's computational complexity and storage requirements increase exponentially with the number of satellites and users, making it impractical for large-scale satellite networks. 

Another approach is modifying the network structure on the base of standard 5G network \cite{stateless_mobile,9755268,9367419,10121544}. This includes introducing new NFs and leveraging advanced 5G handover procedures. However, most of these methods do not specifically prioritize handover latency optimization or provide sufficient improvements in reducing latency. A recent work \cite{stateless_mobile} introduced stateless network elements on the satellite side, allowing users to store the information required by core network, thus avoiding interactions with core network during handover switches. 
However, this work primarily focused on the core network and neglected RAN, which could lead to an underestimation of the overall latency in handover procedure. 

Our proposed scheme involves predicting the network topology (i.e., the user's access satellite). However, there is no need to compute the global topology; instead, each UPF calculate the users it is responsible for respectively, significantly reducing the computational load. 
On the other hand, we modifying the handover procedure. By completely eliminating the interaction with core network during handover, we achieve a substantial reduction in latency, surpassing existing methods.  Our solution takes both core network and RAN into consideration, and we conducted comprehensive system-level experiments on a latency platform. This ensures that experimental results are close to real-world scenarios. 

\section{Discussion and limitation}
\textbf{Access Satellite Selection}:
In this paper, we have evaluated the performance of different kinds of access satellite selection strategies. We find that the consistent strategy can considerably reduce the number of handover while the flexible strategy can provide higher-quality service yet result in more handover. It still remains an open problem whether these two access satellite selection strategies or another option is suitable for the mobile satellite network, which is beyond the scope of the research work in this paper. We believe the access satellite selection scheme is highly related to the applications and leave it for future work.     




\textbf{Access Satellite of Ground Station}: In this paper, we have only discussed the update process of access satellites. However, changes can also occur for the access satellites corresponding to ground stations. Due to the relatively small number and fixed positions of ground stations, satellites can adapt to these changes through updates in their own routing strategies, without the need for reflection in the mobile network's signaling procedures.
%
\section{Conclusion}
Frequent handover resulting from the fast traveling speed of LEO satellites is a significant challenge in mobile satellite networks. In this paper, we added{have proposed} a novel handover design to reduce handover latency. Basing on predicting user access satellites using ephemeris, \added{the propoesd handover scheme has avoided} avoid communication with core network. \added{Furthermore, we} constructed a mobile satellite network prototype using Open5GS and UERANSIM and conducted experiments to evaluate the performance of the proposed handover scheme. The results show that our approach \added{has} achieved a remarkable 10$\times$ reduction in latency compared to the \added{3GPP NTN standard} handover scheme and two other existing approaches.


\bibliographystyle{IEEEtran}
\bibliography{ref.bib}

\end{document}